\documentclass[12pt,preprint]{aastex}

\def\teff{T$_{\rm eff}$}
\def\logg{log g}

\begin{document}

\title{Chemical Abundances of the Leo II Dwarf Galaxy}

\author{Matthew D. Shetrone}
\affil{University of Texas, McDonald Observatory, HC75 Box 1337-McD,
    Fort Davis, TX, 79734}
\author{Michael H. Siegel}
\affil{University of Texas -- McDonald Observatory, Austin, TX, 78712}
\author{David O. Cook}
\affil{Department of Astronomy, University of Minnesota, 116 Church Street SE, Minneapolis, MN 55455}
\author{Tammy Bosler\footnote{AAAS Science \& Technology Policy Fellow}}
\affil{Division of Astronomical Sciences, National Science Foundation}

\begin{abstract}
We use previously-published moderate-resolution spectra in combination with stellar atmosphere
models to derive the first measured chemical abundance ratios 
in the Leo II dSph galaxy.  We find
that for spectra with SNR $> 24$, we are able to measure abundances from weak 
Ti, Fe and Mg lines located near the calcium infrared triplet (CaT). We also quantify and discuss discrepancies between the 
metallicities measured from Fe\,I lines and those estimated from the CaT features.
We find that while the most
metal-poor ([Fe/H] $<-2.0$]) Leo II stars have Ca and Ti abundance ratios similar to those
of Galactic globular clusters, the more metal-rich stars show a gradual decline of Ti, Mg and Ca abundance
ratio with increasing metallicity.  Finding these trends in this distant and apparently
dynamically stable dSph galaxy supports the hypothesis 
that the slow chemical enrichment histories of the dSph galaxies is universal, 
independent of any interaction with the Milky Way.  Combining our spectroscopic abundances with
published broadband photometry and updated isochrones, we are able to approximate stellar ages
for our bright RGB stars
to a relative precision of 2-3 Gyr.  While the derived age-metallicity relationship of Leo II
hints at some amount of slow enrichment, the data are still statistically consistent with no
enrichment over the history of Leo II.
\end{abstract}


\section{Introduction}
Dwarf galaxies are the most common type of galaxy in the universe and are thought to be
the progenitors of larger structures.  Therefore, a clear understanding of dwarf galaxy evolution 
is the first step toward a larger understanding of general galaxy evolution.  Fortunately,
dwarf galaxies are also comparatively simple systems that can
be effectively modeled and simulated. In particular, the satellite galaxies
of the Milky Way, by virtue of being nearby, compact and situated in the sparser regions
of the Galactic halo, have proven to be
exceptional laboratories for studying models of stellar and chemical evolution (see, e.g. Mateo 1998).

A useful tool for studying the dwarf galaxies is the comparison of their chemical
properties to those of the well-studied Milky Way globular clusters.  Some globular
clusters show evidence of multiple stellar populations (Omega Centauri (Lee et al. 1999); NGC~1851 (Milone et al. 2008),
NGC~2808 (Piotto et al. 2007)).  Others, particularly those associated with dSph galaxies or tidal
streams, show
peculiar chemical enrichment histories and/or young ages (see, e.g., Letarte et al. 2006).
However, most {\it inner halo} Milky Way globular clusters formed their
stars a Hubble time ago in a single burst (Marin-Franch et al. 2008).  Since the 
early Universe was dominated by massive stars, it is likely that only core collapse supernovae 
(type II SN) were able to enrich the primordial interstellar medium (ISM)
before the globular clusters formed.  These events would have enriched the ISM
with both Fe and $\alpha$-elements, resulting in  relatively high $\alpha$-element to 
iron ratios [$\alpha$/Fe].  By contrast, objects that formed later, after
type Ia SN had time to enrich the ISM with more Fe, would have lower
relative $\alpha$-element ratios.

Most dwarf galaxies have compound populations, having undergone slow sporadic star
formation over their lifetimes (see review in Mateo 1998).  While their earliest populations
were formed in the aftermath of the the first SNII events, their later populations
benefited from ongoing SNIa.  The result is a steep decline in $\alpha$-element abundance
ratios with decreasing age/increasing metallicity.  The most metal-poor stars have
$\alpha$-element ratios similar to globular clusters, while 
the more metal-rich populations are deficient in $\alpha$-elements
(Shetrone et al. 1998, 2001, 2003; Tolstoy et al. 2003; Venn et al. 2004;
Smith et al. 2006; Helmi et al. 2006).

The chemistry of dwarf Spheroidal (dSph) stars is of particular importance in untangling the hierarchical
formation of the Milky Way.  CDM cosmological models predict that galaxies like the Milky
Way have grown through the merging of low-mass systems with perhaps some low-level merging continuing
at the present time.
Among the current retinue of dSph satellite galaxies, the Sagittarius
dSph shows clear signs of having been torn apart by the Milky Way
potential (Majewski et al. 2003) and Ursa Minor, Leo I, Coma Berenices and Ursa Major II
also hint at a violent dynamical past (Palma et al. 2003; Sohn et al. 2007; Simon \& Geha 2007).

However, while the most metal-poor stars of the dSph galaxies are chemically similar to the most
metal-poor stars of the Galactic halo, the enrichment of SNIa kicks in at lower Fe abundances 
in the dSph galaxies, resulting in consistent deficits in the $\alpha$-element ratios of dSph
stars compared to halo stars of similar [Fe/H] (Venn et al. 2004).
This would indicate that the Galactic halo formed from objects that
had more rapid and therefore more SNII-dominated chemical evolution and that objects like the current
dSph galaxies have contributed little to the halo.

This argument is, however, undermined by recent studies indicating that
the stars the Sagittarius dSph is contributing to the halo differ from those in the residual core 
(Bellazzini et al. 2006; Chou et al. 2007; Siegel et al. 2007).  If the dSph galaxies
are the remnants of much larger objects that have been disrupted over a Hubble time, 
the {\it present} stellar populations of the dSph galaxies may not reflect the {\it initial} stellar 
populations they would have preferentially contributed to the Galactic halo.
Moreover, if the stellar populations of the dSph galaxies have not evolved in isolation, their
chemical abundances may have been affected by continual dynamical harassment.

These arguments over dSph stellar chemistry could be clarified by the examination of more distant 
dwarf galaxies
that have not been perturbed by the Milky Way potential.  If an isolated dSph were shown
to have the same chemical abundance patterns as the relatively nearby dSphs, such
a discovery would (a) strengthen the argument that the lack of halo stars with ``dSph-like" abundance
patterns suggests that the vast majority of the halo formed  
very quickly and that the dSphs have contributed very few stars since this 
initial construction phase; and
(b) show that the $\alpha$-element deficits have
nothing to do with dynamical stirring and are entirely the product 
of dSph stellar and chemical evolution.

In an effort to address these issues, we now expand 
the study of chemical abundances
to the Leo II dSph galaxy.
Leo II is the second most distant dSph galaxy assumed to be orbiting 
the Milky Way (218 kpc, Bellazzini
et al. 2005; Siegel et al. 2008, hereafter S08).  It is predominantly
metal-poor (Hodge 1982; Demers \& Harris 1983; 
Aaronson et al. 1983; Azzopardi et al. 1985; Suntzeff et al. 1986;
Lee 1995; Demers \& Irwin 1993; Koch et al. 2007a,b; Bosler et al. 2007, hereafter BSS07) and 
dominated by intermediate age populations (Mighell \& Rich 1996; Dolphin
et al. 2005; Bellazzini et al. 2005; Gullieuszik et al. 2008).  It has a handful of carbon stars (Azzopardi et al. 1985), a 
rich population of 
RR Lyrae variable stars (Swope 1967, 1968; Siegel \& Majewski 2000), a
high central velocity dispersion (Vogt et al. 1995; Koch et al. 2007b; BSS07; S08) and
an extended stellar distribution (Coleman et al. 2007; S08).  At present, however, the indications are that
is had undergone little, if any, interaction with the Galactic potential (Koch et. al 2007b; 
Coleman et al. 2007; Walker et al. 2007; S08) and its modest radial velocity (+26.2 km s$^{-1}$, S08) would
be consistent with a circular orbit that does not bring it close to the Milky Way.

The BSS07 study cited above analyzed medium resolution NIR spectra of a large
number of Leo II RGB stars in an effort to measure radial velocities and metallicities.
These spectra are dominated by the Calcium II Infrared Triplet (CaT) features
but also have numerous weak Fe, Ti, and Mg lines that have yet to be exploited.
In this paper, we show that these weak lines can be used
to measure metallicity and abundance ratios for
stars of sufficient S/N through the use of synthetic model spectra.  This
method has allowed us to study the chemical evolution of Leo II, to
constrain the relationship between the equivalent width of the CaT lines
and the [Fe/H] abundances as measured from Fe\,I lines and to make a
preliminary examination of the age-metallicity relationship (AMR) of the
brightest Leo II RGB stars.

\S2 of this paper discusses the additional reduction and analysis we have applied to the 
BSS07 data. \S3 
compares the abundances ratios measured in Leo II to those measured in the globular clusters to
provide insight into the chemical evolution of Leo II. \S4 examines the inferred ages of our bright Leo II
stars while \S5 summarizes our findings.

\section{Spectroscopic Reduction and Analysis Techniques}

A full description of the observations and data reduction can 
be found in BSS07 but we give a brief summary of the observations here.
Low-dispersion spectra of red giants in the Leo II dSph
were taken using the Keck I 10-m telescope and LRIS (Oke et al. 1995) 
in 2002 and 2003.   The LRIS was configured with the 
1200 l/mm grating blazed at 7500 \AA which gives a dispersion of 0.62 \AA 
per pixel and resolution of 1.55 \AA (R $\sim$ 10 000).   Twenty to thirty
stars were observed on each slit mask creating a total sample of 74 stars
with SNR between 10 and 44.

Traditional equivalent width analysis of these spectra proved impossible due to the
heavy blending of many lines and the difficulty in setting the continuum measure.  However,
these problems can be overcome by comparison of the observed spectra to 
synthetic spectra created by the 2007 version of the LTE line-analysis 
code MOOG (Sneden 1973).

The initial line list was obtained from the R. L. Kurucz CD-ROM 23. 
In order to further calibrate this line list we compared the list with the 
nearby red giant star, Arcturus (Hinkle et al. 2000). 
We adopted atmospheric parameters for Arcturus
(\teff = 4280, \logg = 1.55, $v_t$ = 1.61 km s$^{-1}$, 
[Fe/H] = -0.50, [Ti/Fe] = +0.26, 
[Mg/Fe] = +0.39 and [Ca/Fe] = +0.21)   
by combining the results of McWilliam (1990), Fulbright et al. (2006,2007) 
and Koch \& McWilliam (2008).  Since we are performing a differential analysis, we chose
to adjust the gf values in our synthetic spectrum only for lines that
were significantly discrepant from the catalog Arcturus spectrum, rather than adjust all 8345 lines.

We smoothed the model spectra with a Gaussian kernel of FWHM equivalent to the BSS07 spectral
resolution of 1.55 \AA.  Initial metallicity estimates for globular cluster and Leo II stars
were taken from BSS07.  Effective temperatures were
calculated from interpolation of the temperature-color-metallicity
relationship of Ramirez \& Melendez (2005) and the $BV$ photometry of S08, the latter
allowing the Leo II and cluster data to be analyzed using the same photometric system.

Surface gravities were calculated from the temperatures and bolometric-corrected
magnitudes assuming a mass of 0.8 $M_{\circ}$.
We used distance moduli of 15.07, 13.76, 15.59, 15.19 and 21.61 for M3, M107, NGC~1094, NGC~4590
and Leo II,
respectively.\footnote{We find in \S4 that Leo II's distance modulus is likely closer
to 21.7.  Correcting the Leo II distance modulus would change the inferred $log g$ values
by approximately 0.04, an insignificant correction.}  The microturbulence was 
estimated as $v_t =  -0.41 * log g + 2.15$, in accordance
with recent analyzes of 
globular cluster giants near the tip of the giant branch 
(e.g. Sneden et al. 2004, Lee, Carney \& Habgood 2005, Ivans et al. 2001, Cohen
\& Melendez 2005).  Microturbulent velocities ranged from 1.4 km s$^{-1}$ to 2.0 km s$^{-1}$  
in the globular cluster sample and from 1.6 to 2.0 km s$^{-1}$ in the Leo II sample.

These parameters were fed into MOOG to produce preliminary synthetic
spectra.  We then adjusted the [Fe/H] metallicity
of the atmospheres until the Fe lines matched the observed spectrum.
A weighted mean of all the iron lines was used to determine the 
best [Fe/H] value for each star.

The model was then set to this metallicity and the abundance 
ratios were determined by holding all elements constant while varying the 
desired $\alpha$-element.  The best fit was determined by the 
output difference between the MOOG synthetic spectra and observed spectra.
We found that we were able to effectively measure abundances for stars 
with SNR $> 24$ per pixel.

The setting of the continuum was critical for a proper analysis of 
the weak lines in these 
medium resolution spectra.  To refine the continuum levels we created a 
synthetic spectrum for
each star, divided it into the observed spectrum and fit 
the residual with a high
order spline.  This fitted spline was then divided into the observed 
spectrum to remove high order
terms in the continuum placement.  If our modeled spectrum changed 
significantly, e.g. when a 
new overall metallicity change was required, we repeated this procedure 
for the new model.

We attempted to determine abundance ratios for all of the moderate lines 
in the spectra.  These lines include Fe, Ca, Na, Mg, and Ti.   
Figure 1 compares one of our highest SNR spectra to several synthetic 
spectra with varied Fe and Ti
abundances.  There are a large number of lines in this region with varying
strength that can be used to determine abundances.  For a more typical SNR spectrum, most
of these lines are not strong enough for abundance analysis.  We identified
sufficient Ti and Fe lines in all of our bright stars but the Na lines and all 
but one of the Mg lines were absent in 
most of the spectra.  Thus we limit our analysis to the stronger Fe, Ca, Mg 
and Ti lines listed in Table 1.

The only Ca lines in the observed spectral region are the extremely
strong Ca triplet lines.  The CaT lines are difficult to model with synthetic
spectra (hence the more common use of globular cluster-calibrated equivalent width
analysis).
At a constant Ca abundance, the strength of the CaT lines decreases
as the electron pressure increases due to changes in the 
continuous opacities.  For cool stellar atmospheres, the 
main sources of electrons are Mg, Fe, Si, Ca, Na and Al, but which element 
contributes the most depends in the effective temperature of the star
and which layer of the atmosphere is surveyed.  For example, for 
a \teff = 4200 K star at 0.01 optical depth, the main sources of electrons are Mg and 
Fe; while for a \teff = 4000 K star at 0.01 optical depth,
the main sources of electrons are Ca, Mg, Na and Al.  
Thus, as the $\alpha$-abundance declines, the continuous
opacity changes to counteract the Ca abundance decline.  
For example, a 0.5 dex change in the electron contributors will increase the CaT 
line strength enough to mimic a Ca abundance increase of 0.38 and 
0.22 dex for stars of \teff=4000, \logg = 0.7, [M/H] = -1.1 
and \teff=4200, \logg = 0.7, [M/H] = -1.9, respectively.   
This means that as [M/H] rises, any decline in the [Ca/Fe] ratio
would be muted in the measured CaT lines.  As a further complication, 
the CaT lines become less sensitive to the Ca abundance 
as the lines strengthen.

For the globular cluster sample, the Plez model atmospheres with 
[$\alpha$/Fe]=+0.4 enhanced abundance ratios were appropriate
and there was no need to change the model.  However, our initial analysis
showed that the Ti and Mg ratios of the Leo II stars decline with increasing metallicity (\S3).
This resulted in stars with low $\alpha$-abundances having erroneously high Ca abundances.
To derive accurate Ca abundance, we were therefore required to iterate.  We first 
determined the Fe abundance for each star and
then adjusted the model appropriately.  We next determined the Ti and Mg 
abundances, from which
we calculated a preliminary $\alpha$-abundance. Finally, we adjusted the 
model abundance based on the average $\alpha$-abundance at each star's [Fe/H].   
The result was that for the more metal-rich Leo II stars, we used model
atmospheres that were more metal-poor than the Fe abundance would imply, thus
compensating for the lack of electrons owing to the lower $\alpha$-abundances.

Measurement errors were determined for each fitted feature by adjusting 
the abundances up and down until the residuals of the fit were larger 
than the surrounding continuum regions, ie. larger than the SNR.   The measurement
errors were then propagated to yield a single measurement error which was
added in quadrature to the abundance errors from the 
modeling uncertainties -- $\pm100$ K for \teff, $\pm0.2$ dex for \logg, 
$\pm0.25$ for v$_t$ and $\pm0.2$ dex for [M/H].  Typically the 
measurement errors dominated over the modeling errors particularly for 
[Mg/Fe] which was derived from a single spectral feature.

Tables 2 and 3 list the globular cluster and Leo II stars from BSS07 for
which we have been able to determine abundances while tables 4 and 5 show the abundances
derived from the MOOG software.
We have recalculated the signal to noise ratio (SNR) from the 
residual of the best fit synthetic spectra to the observed spectra
and list that SNR in tables 2 and 3.

\section{Spectroscopic Results}

Figure 2 shows the photometrically derived effective temperatures and surface gravities for the Leo II sample.
There is a clear RGB sequence, corresponding to the photometric sequence upon which the 
temperatures and gravities are based.  The two outlier stars -- 195 and 166 -- 
are metal-poor radial velocity members. They are, however, 
according to S08, well-removed from the dominant Leo II RGB locus in color-magnitude space although both
have Washington+DDO51 photometry consistent with giant stars.

It is possible that these are interloping giants
in the field, possibly from the metal-poor debris stream of the Sagittarius dSph galaxy passing 
in front of Leo II (S08).  Alternatively, they could represent a young, metal-poor population in LeoII
(\S4).  Given their outlier status, we have chosen to remove these two stars from
the analysis of mean trends in the Leo II sample.

Figure 3 compares our derived [Fe/H] metallicities against those
derived from the CaT lines by BSS07.  
On average, the CaT-based metallicities are 0.17 dex
higher than the our model [Fe/H] metallicities with an RMS of 0.18 dex.
The abundances are correlated
but there is a positive residual which shows increasing significance with 
decreasing metallicity.   

The offset is likely due to several factors.  First, all of our
Fe lines are Fe\,I lines and it is possible that metal-poor giants
overionize Fe (Thevenin \& Idiart 1999; Asplund \& Garcia-Perez 2001).
In M3, for example, Sneden et al. (2004) find an underabundance factor
of -0.13 in the Fe\,I lines and they adopt the Fe\,II abundances
as the true metallicity.  
In M68 Lee, Carney \& Habgood (2005) find an underabundance factor of 
-0.37 in the Fe\,I lines and also adopt the Fe\,II abundances as the true metallicity.  
However, not all authors find these
overionization factors.  Cohen \& Melendez (2005) find that the Fe\,I and
Fe\,II abundances agree to within uncertainties.  Kraft \& Ivans (2003) find small
and variable overionization in metal-poor globular clusters but of much 
smaller magnitude than Thevenin \& Idiart (1999) predict.
In light of the disagreement, we have chosen not to make any
correction for overionization but will discuss the potential impact in later sections.

A second potential source of the discrepancy between the [M/H]$_{CaT}$
and our derived [Fe/H] comes from the calibration of the CaT metallicity 
scale, which in BSS07 is based on the Carretta \& Gratton (1997) metallicity scale.  
The choice of calibration system can significantly impact the 
zero point of the metallicity.   Battaglia et al. (2008), for example, 
find a 0.1 dex 
difference between the Carretta \& Gratton (1997) 
metallicity scale and their high resolution analysis
over the range of metallicity spanned by Leo II's RGB stars.

We can explore the issue of the metallicity zero point by comparing
the Fe\,I metallicities we derive for the BSS07 globular clusters to the
high resolution analyzes listed in Pritzl, Venn \& Irwin (2005).  The average abundances
we derive
for NGC~1904, NGC~4590 and M3 are -0.14 ($\sigma = 0.22$) dex more metal-poor 
than those listed in Pritzl et al..  However, their NGC~1904 abundances
come solely from the study of Gratton \& Ortolani (1989).  Pritzl et  al. find
that the Gratton \& Ortolani abundances are typically 0.16 more metal-rich than
more recent analyzes.  Correcting the
NGC~1904 abundance down by 0.16 dex, we find our calculated
Fe I abundances for NGC~1904, NGC~4590 and M3 are -0.08 (0.14) dex more metal-poor
than the Pritzl et al. values -- a zero point offset within the
error of the weighted mean abundances listed in Table 4.

Figures 4 and 5 show the derived [Ti/Fe] and [Mg/Fe] abundance 
ratios, respectively, for both the Leo II and 
globular cluster sample.  The Leo II and globular cluster star [Ti/Fe]
abundances overlap at the poorest metallicities but depart dramatically 
at higher metallicities.  The Leo II [Mg/Fe] ratios
are offset from the globular cluster sample and show a pronounced
decline with increasing metallicity.  The large error-bars on the [Mg/Fe]
limit the usefulness of the individual measurements but confirm the
same trend seen in the [Ti/Fe] abundance ratios.

We note that the [Ti/Fe]
ratios in M107 are as high if not higher than those found in the 
more metal-poor globular clusters.   Such large [Ti/Fe] abundance 
ratios are not unprecedented.  Sneden et al. (1994) found that the relatively 
metal-rich globular cluster M71 had high Ti abundance ratios 
([Ti/Fe] = +0.5), although
Ramirez \& Cohen (2002) found "normal'' Ti ratios M71.

Figure 6 shows the [Ca\,II/Fe\,I] abundance ratios for the globular cluster and Leo II samples.
There is a slight decline 
in the [Ca\,II/Fe\,I] ratio with increasing metallicity among the
globular cluster sample.  The Leo II Ca abundance ratios are
offset from the globular cluster sample and the slope of the best fit line is 
slightly steeper than that of the cluster sample.  This is very similar to 
the trends in Mg abundances seen in Figure 5.   

For M3, NGC~4590 and NGC~1904, we compare our abundance ratios to
the compilation
of Pritzl, Venn \& Irwin (2005).  Our [Ti/Fe] ratios
are $0.09\pm0.05$ dex larger,  
our [Mg/Fe] ratios are $0.01\pm0.05$ dex larger, and
our [Ca/Fe] ratios are $0.24\pm0.13$ dex larger.   
The only significant
zero point is in our [Ca\,II/Fe\,I] abundances.  
This is not entirely unexpected.  Most analyzes compare 
ionized species to ionized
species and neutral species to neutral species.  We would obviously
prefer to make a similar comparison, but the previously noted lack of Fe\,II
lines prevents this.  Additionally, as noted above, the literature is divided on 
the over-ionization correction need to converted from Fe\,I to Fe\,II abundances.
Such corrections are likely to be metallicity dependent and could be
the cause of the slope in the globular cluster sample.
Because we do not make a correction, the absolute value of our [Ca\,II/Fe\,I] abundances
is uncertain.  However, because our analysis is concerned with {\it relative}
differences between the clusters and Leo II, such a zero point correction
will have little impact on our conclusions.

Figure 7 show the [Ca/H] ratios derived from our analysis 
compared to the [Ca/H] ratios derived from the CaT by
BSS07.  The overlap of the cluster and Leo II sample in this figure suggests that
the corrections we made for the more metal-rich low-$\alpha$ stars are 
consistent with Ca abundances derived with the CaT methodology.
The slopes in these samples are less steep than 
those of Figure 3 -- as would be expected since the CaT lines should be 
sensitive to the overall Ca abundance.  However, there is a residual
slope between the two [Ca/H] scales, which remains unexplained.

Figure 8 shows the mean of the [Ca/Fe], [Ti/Fe] and [Mg/Fe] ratios.  The
fit to the data with the lowest formal reduced chi square
is a second order Legendre polynomial which exhibits
a steep decline in the $\alpha$-abundance ratios of the Leo II
stars with metallicities above 
[Fe/H]$\sim$-2.0.  This is consistent with
what is seen in other dSph galaxies (see references in \S1) and is likely 
due to slower chemical evolution and concordant increased
contribution of SNIa to Leo II.  However, the data would also be consistent with a
step function near [Fe/H]$\sim$-1.8 in which the metal-poor stars have globular cluster
like $\alpha$-abundances while the more metal-rich stars are underabundant in
$\alpha$ elements.

\section{Age Distribution of Leo II Stars}

The slow enrichment implied by our abundance analysis suggests that Leo II should
have an age-metallicity relationship (AMR) in which the more metal-rich stars are younger
than the more metal-poor stars and have therefore been subject to more enrichment
by SNIa.  An AMR would not be immediately obvious in color-magnitude
diagrams due the photometric degeneracy of age and metallicity.  However, our precise
spectroscopic survey affords the ability to remove abundance from the equation.  We can now
combine precise broadband photometry with modern isochrones to approximate individual stellar ages.
This method has had some success in untangling the stellar populations of 
the Omega Centauri globular cluster, for example (Hilker et al. 2004; Sollima et al. 2005)
and was recently applied to the Leo II dSph galaxy by Koch et al. (2007a).

The Koch et al. study
found that Leo II has undergone steady star formation with little enrichment, starting 9 Gyr ago and continuing up until
2 Gyr ago.  More recently-formed stars appear to be somewhat enriched.
Our analysis uses similar data but with the advantage that
we have measured $\alpha$-abundances, rather than assumed abundances.

Because our stars are near the tip of the RGB, rather than the more age-sensitive turnoff, 
any analysis will necessarily be imprecise (Koch et al. estimate uncertainties of 40\% using
these techniques).  The ideal stars for this analysis are those near the age-sensitive main sequence
turnoff (MSTO).  However, given that Leo II's
MSTO is well beyond the spectroscopic range of even the largest ground-based telescopes,
the RGB is our best option for probing Leo II's age distribution.

Figure 9 demonstrates the analysis.  Photometry is taken from S08, which has a precision
of 0.01 mag in all passbands at the tip of the red giant branch (TRGB).  $VI$ colors were calculated
by converting the Washington measures in S08 using transformations in Majewski et al. (2000).
Isochrones are taken from the Dartmouth Stellar Evolution
Program (Dotter et al. 2007) and shifted to a distance modulus of $(m-M)=21.70$ and reddening
of $E(B-V)$=0.02, in
accordance with S08's TRGB distance derived from the same data.

Cursory examination of the 
individual panels, broken
down by passband and metallicity bin, shows  no obvious AMR in Leo II.
To make a more precise estimate, we interpolated the isochrones in 0.01 magnitude intervals and,
with metallicity fixed at the spectroscopic value and [$\alpha/Fe$] fixed to the mean [$\alpha/Fe]$ at
each star's [Fe/H], we found the isochrone nearest to each star's
photometric position.  We estimated age uncertainties by offsetting the [Fe/H] of each star 
by its uncertainty (which dominates over S08's small photometric uncertainties).

The left panels of Figure 10 show the resultant age-metallicity distribution.
Although the uncertainties are 
significant (typically 2-3 Gyr), some interesting trends can be seen.  The mean age of the stars is $\sim9$ Gyr but has a spread
significantly greater than the uncertainty, implying ongoing low-level star formation.  The younger stars appear
to be slightly more metal-rich and indeed, an AMR that shows slight enrichment with age -- either gradually or step-wise -- 
would be more consistent with the data than a flat trend.  However, we can not statistically rule out a flat AMR, which would have a $\chi^2$ per
degree of freedom of 0.9.
In fact, our derived AMR is very similar to of Koch et al. 2007a.

It should be noted that the exploration of the RGB stars has number of complications
that could affect the inferred AMR. In particular:

$\bullet$  Leo II is known to have prominent asymptotic giant branch (AGB; Mighell \& Rich 1996; Gullieuszik et al. 2008).
Stars on the AGB will
overlap young RGB isochrones and perhaps 10\% of our stars could be AGB stars.  The most likely
candidates would be the two apparently young metal-poor stars that are clear outliers in the bottom
panels of Figure 9 and both panels of Figure 10 (note that these are different stars than
the two outliers rejected in Figures 2-8, which were excluded {\it a priori} from the age analysis).  We attempted to measure ages for these stars using the Padova
isochrones (Marigo et al. 2008), which include an AGB.  However, these stars are too blue for all but the youngest AGB isochrones
(at these magnitudes,
the RGB and AGB are separated by only a few .01 mag).  Removing these outlier stars from the sample, however, increases the contrast
between the younger and older stars of Leo II, making enrichment more tenable.  But a flat AMR
would still be within the uncertainties ($\chi^2 = 0.9$).

$\bullet$ Any age analysis will be sensitive to the distance modulus of Leo II.  Adjusting the distance
of Leo II shifts the absolute ages, but not the relative ages.  The exception is at short distance
moduli, where the stars are too faint to correspond to anything but the maximum age in our
isochrone set (15 Gyr).  If we assume that Leo II's stars must be younger than 15 Gyr, then Leo II
must be at a distance of at least $(m-M)=21.7$.  Longer distances result in younger absolute ages
but have little effect on the inferred AMR.

$\bullet$ The overionization correction mentioned in \S2 could alter the AMR by shifting
the sample to a more metal-rich distribution. We applied a 0.2 dex correction (right panels of Figure
10) and find that the AMR tightens somewhat, producing a sharper peak in the age distribution at around 6-7 Gyr.  However, a
flat AMR would still be within the uncertainties ($\chi^2 = 0.9$), even if the two outlier stars are removed.

$\bullet$ Our SNR-limited sample is unable to probe the fainter more metal-rich stars in Leo II.
Improved spectra of these stars would extend the AMR over a greater range
of metallicity, providing better constraint of the enrichment and allowing a comparison
to toy models of enrichment.  In particular, it would determine if the apparent rise in abundance for the younger
stars in Figure 10 is merely statistical scatter or is the indicator of a larger trend.

Even with these caveats, the age-metallicity distribution of Leo II stars, while giving tantalizing hints of enrichment,
remain statistically consistent with no enrichment.  A flat enrichment would be unexpected for Leo II, since we 
{\it do} find that the more metal-rich stars have lower $\alpha$-element abundances, suggesting gradual enrichment by SNIa over 
the course of several Gyr. 

In the end, while this exercise demonstrates that precise photometry and spectroscopy
can be used to probe the AMR of objects out to 200 kpc, the enrichment of Leo II may be too subtle to be detected in our data.
Spectroscopic examination of the more numerous and age-sensitive turnoff stars may be the only way to tease out the
AMR of this distant and enigmatic galaxy.

\section{Discussion and Conclusions}

Our analysis of the weaker spectral features in the BSS07 spectra provides 
critical information about not only the Leo II dSph galaxy 
but other spectroscopic studies as well.

$\bullet$ We find the Fe-line metallicities disagree with CaT-based 
metallicities, although it is likely this is a result of the
overionization of Fe at low
metallicity.  We are unable to analyze the fainter more metal-rich 
stars to confirm the high-metallicity tail identified by BSS07.

$\bullet$ We confirm that Leo II, like the other low-luminosity dSph 
galaxies, has globular cluster-like $\alpha$-abundances in its most metal-poor 
stars but shows a declining $\alpha$-abundance at higher metallicities, likely
due to the increasing influence of SNIa.

$\bullet$ Leo II's AMR gives tantalizing hints of slow enrichment.  However, a flat
AMR with no enrichment can not be ruled out with the present data, given
the inherent uncertainties in analyzing only the brightest stars in the galaxy. 
Nevertheless, this demonstrates that precise spectroscopic and photometric data can be used
to get a general picture of the enrichment history of distant dSph galaxies.

Leo II's particular influence on models of dwarf galaxy evolution lies 
in its difference from the other
dSph galaxies.  As noted in \S1, Leo II appears to have had little interaction
with the Milky Way.  This would indicate that
the $\alpha$-abundance patterns 
found in the dSph galaxies are a universal
feature of dwarf galaxies and have no relation
to any dynamical interaction with a larger parent galaxy.  
It also strengthens the contention
that while objects like the present retinue of dSphs
have contributed somewhat to the construction of the Milky Way -- given the similarities
between their most metal-poor stars and some chemically peculiar stars within the halo --
the stark chemical dissimilarities between the bulk of the dSph 
and field halo stars indicates that this contribution has been small.

Further investigation is needed into the zero point issues intrinsic in our analysis
so that we may more confidently translate measured Fe\,I abundances into intrinsic Fe\,II abundances.
This could be accomplished by looking at existing high 
SNR CaT spectra of globular clusters, especially those known 
to be deficient in $\alpha$-elements or those known or suspected
to have been stripped from larger dSph galaxies.
Once the zero point issues are understood, the technique we have unveiled in
this paper can be applied to the large CaT spectral datasets in the 
literature to derive $\alpha$-abundance measures for thousands of dSph and field
halo stars.  Better and easier chemical fingerprinting of distant RGB
stars has the potential to finally 
unravel the hierarchical history of the Milky Way.

Moreover, this technique expands the range of current and future investigations into
dSph chemistry.  It shows that current telescopes can probe dSph chemistry out to at least
200 kpc and indicates that next-generation telescopes will open up 
investigation into even more isolated dwarfs such as the Phoenix dIrr/dSph (Tobolewski
et al., in prep.).  This will expand the study of galactic chemical evolution into an even
more diverse array of galactic environments.  At this time, however, 
the existing models of dwarf galaxy chemical evolution appear to be 
valid out to at least 200 kpc from the Galaxy.

\acknowledgments

This project was completed during
the McDonald Observatory REU and was supported under NSF AST-0649128. 
MHS and MDS were supported by NSF grant AST-0306884.

\clearpage

\begin{deluxetable}{lrrrrrrr}
\tablewidth{0 pt}
\tablecaption{Main lines used in Abundance Analysis }
\tablehead{
\colhead{Element } &
\colhead{Wavelength } &
\colhead{eV } &
\colhead{log gf } }
\startdata
   Ti\,I  &   8378.25  &  3.72  &  -2.30  \\
   Ti\,I  &   8382.53  &  0.82  &  -1.63  \\
   Ti\,I  &   8435.65  &  0.84  &  -1.30  \\
   Fe\,I  &   8220.38  &  4.32  &   0.20  \\
   Fe\,I  &   8327.06  &  2.20  &  -1.64  \\
   Fe\,I  &   8387.77  &  2.18  &  -1.60  \\
   Fe\,I  &   8468.41  &  2.22  &  -2.17  \\
   Fe\,I  &   8514.07  &  2.20  &  -2.25  \\
   Fe\,I  &   8611.80  &  2.85  &  -2.15  \\
   Fe\,I  &   8674.75  &  2.83  &  -2.00  \\
   Fe\,I  &   8688.63  &  2.18  &  -1.21  \\
   Mg\,I  &   8806.76  &  4.35  &  -0.13  \\
   Ca II  &   8498.02  &  1.69  &  -1.31  \\
   Ca II  &   8542.09  &  1.70  &  -0.36  \\
   Ca II  &   8662.14  &  1.69  &  -0.62  \\
\enddata
\end{deluxetable}

\clearpage

\begin{deluxetable}{lrrrrrr}
\tablewidth{0 pt}
\tablecaption{Cluster Sample take from Bosler Survey }
\tablehead{
\colhead{Stars } &
\colhead{V (mag) } &
\colhead{B-V } &
\colhead{SNR } &
\colhead{\teff}&
\colhead{\logg }&
\colhead{v$_t$ } }
\startdata
\multicolumn{7}{c} {NGC1904} \\
\hline
   241  &13.61     &  1.11   &  37.6    &  4419  &    0.94  &  1.77     \\
   131  &13.02     &  1.20   &  34.9    &  4298  &    0.85  &  1.80     \\
   223  &13.19     &  1.16   &  34.5    &  4350   &    0.72  &  1.85     \\
   160  &14.06     &  1.52   &  31.2    &  3969  &    0.31  &  2.02     \\
   294  &14.22     &  1.00   &  22.3    &  4586  &    1.29  &  1.62     \\
   181  &13.95     &  1.25   &  21.3    &  4236  &    0.94  &  1.76     \\
\hline
\hline
\multicolumn{7}{c} {NGC4590} \\       
\hline
   HI82 &12.59     &   1.33  &  24.6    &  4188   &  0.34  &  1.98 \\
   HI119&13.62     &   0.96  &  21.1    &  4633   &  1.09  &  1.70 \\
   HI239&14.19     &   0.87  &  14.2    &  4788   &  0.89  &  1.79 \\
\hline
\hline
\multicolumn{7}{c} {M3} \\                                        
\hline
     265 &13.26     &    1.30 &  93.0   &     4155    &     0.83   & 1.81 \\
     250 &14.11     &    0.95 &  82.1   &     4627    &     1.50   & 1.54 \\
      640 &13.27     &    1.26 & 77.1   &     4199    &     0.87   & 1.79\\
     589 &12.87     &    1.33  & 76.6   &    4124    &     0.64   & 1.89 \\
     1217 &14.00     &   1.05  & 76.4   &    4470    &     1.36   & 1.59 \\
     885  &13.47     &   1.07 &  75.5   &     4441    &     1.13   & 1.69 \\
     334 &13.24     &    1.20 &  55.3   &     4269    &     0.91   & 1.78 \\
     238 &12.69     &    1.57 &  53.5   &     3918    &     0.35   & 2.01 \\
\hline
\hline
\multicolumn{7}{c} {M107} \\                                      
\hline
   Sl &   14.04  &     1.47  &  56.1    &     4581    &     1.37   & 1.59 \\
   Sf &   13.39  &     1.70  &  49.0    &     4225    &     0.87   & 1.79 \\
   Sr   & 14.66    &   1.29  &  42.2    &     4910    &     1.80   & 1.41 \\
   Sh &   13.84  &     1.61  &  38.9    &     4356    &     1.15   & 1.68 \\
   Ss  &  14.79   &    1.40  &  37.9    &     4703     &     1.74   & 1.43 \\
   Su   &  14.78   &   1.28  &  37.5    &     4930     &     1.86   & 1.39 \\
   S62 &  13.97   &    1.62  &  34.1    &     4341    &     1.19   & 1.66 \\
\hline
\enddata
\end{deluxetable}

\clearpage

\begin{deluxetable}{lrrrcrrr}
\tablewidth{0 pt}
\tablecaption{Leo II Sample take from Bosler Survey }
\tablehead{
\colhead{Stars } &
\colhead{V (mag) } &
\colhead{B-V } &
\colhead{SNR } &
\colhead{[M/H]$_{CaT}$\tablenotemark{a} } &
\colhead{\teff}&
\colhead{\logg}&
\colhead{v$_t$ } }
\startdata
180  &  18.998   &   1.5117   &43.7 &-1.60   &  3981   &    0.33  &   2.02  \\
255  &  19.291   &   1.3114   &40.0 &-1.66   &  4151   &    0.62  &   1.90  \\
271  &  19.375   &   1.3561   &39.5 &-1.48   &  4100   &    0.60  &   1.90  \\
258  &  19.311   &   1.3507   &36.9 &-1.57   &  4110   &    0.59  &   1.91  \\
336  &  19.531   &   1.1669   &36.8 &-2.03   &  4299   &    0.83  &   1.81  \\
195  &  19.026   &   1.0915   &36.2 &-1.60   &  4406   &    0.71  &   1.86  \\
296  &  19.385   &   1.0488   &35.7 &-2.02   &  4438   &    0.87  &   1.79  \\
293  &  19.304   &   1.2481   &35.5 &-1.84   &  4216   &    0.68  &   1.87  \\
166  &  18.827   &   1.1629   &35.3 &-1.52   &  4318   &    0.56  &   1.92  \\
285  &  19.337   &   1.3657   &35.0 &-1.50   &  4091   &    0.58  &   1.91  \\
304  &  19.389   &   1.2711   &34.7 &-1.81   &  4194   &    0.69  &   1.87  \\
254  &  19.259   &   1.3981   &33.6 &-1.43   &  4055   &    0.51  &   1.87  \\
236  &  19.212   &   1.3330   &32.6 &-1.90   &  4144   &    0.58  &   1.91  \\
351  &  19.552   &   1.2835   &32.5 &-1.40   &  4176   &    0.74  &   1.85  \\
333  &  19.495   &   1.2278   &32.4 &-1.45   &  4240   &    0.77  &   1.83  \\
379  &  19.618   &   1.3171   &32.1 &-1.52   &  4141   &    0.74  &   1.85  \\
209  &  19.033   &   1.4749   &31.7 &-1.59   &  4006   &    0.37  &   2.00  \\
256  &  19.332   &   1.2784   &30.4 &-1.54   &  4182   &    0.66  &   1.88  \\
341  &  19.541   &   1.2780   &29.2 &-1.43   &  4182   &    0.74  &   1.85  \\
281  &  19.342   &   1.1127   &28.6 &-2.05   &  4358   &    0.80  &   1.82  \\
282  &  19.329   &   1.4611   &27.6 &-1.52   &  4008   &    0.49  &   1.95  \\
420  &  19.717   &   1.0824   &27.1 &-1.46   &  4430   &    1.00  &   1.74  \\
377  &  19.587   &   1.1853   &25.8 &-1.73   &  4283   &    0.84  &   1.80  \\
260  &  19.268   &   1.2811   &25.7 &-1.57   &  4179   &    0.63  &   1.89  \\
248  &  19.354   &   1.2399   &25.1 &-1.51   &  4225   &    0.70  &   1.86  \\
234  &  19.250   &   1.2238   &24.5 &-1.80   &  4240   &    0.67  &   1.87  \\
230  &  19.194   &   1.4458   &24.1 &-1.56   &  4025   &    0.45  &   1.96  \\
\enddata
\tablenotetext{a}{Based on the BSS07 CaT measurement and the Carretta \& Gratton (1997) 
metallicity scale}
\end{deluxetable}

\clearpage

\begin{deluxetable}{lrrrr}
\tablewidth{0 pt}
\tablecaption{Derived Abundances for Cluster Sample }
\tablehead{
\colhead{Star } &
\colhead{[FeI/H] ($\sigma$) } &
\colhead{[TiI/FeI] ($\sigma$) } &
\colhead{[MgI/FeI] ($\sigma$) } &
\colhead{[CaII/FeI] ($\sigma$) } 
}
\startdata
\multicolumn{5}{c} {NGC1904} \\
     241 &  -1.70  (0.18)  &  0.11 (0.27) &  0.30 (0.38) & 0.70 (0.17) \\
     131 &  -1.89  (0.21)  &  0.70 (0.25) &  0.45 (0.37) & 0.70 (0.20) \\
     223 &  -1.55  (0.18)  &  0.57 (0.22) &  0.50 (0.39) & 0.70 (0.17) \\
     160 &  -2.06  (0.26)  & -0.03 (0.30) &  0.70 (0.59) & 0.35 (0.20) \\
     294 &  -1.62  (0.22)  &  0.65 (0.27) &  0.60 (0.42) & 0.45 (0.17) \\ 
     181 &  -2.02  (0.22)  &  0.46 (0.25) &  0.20 (0.43) & 0.85 (0.24) \\
\tableline
average &   -1.77 (0.08)  & 0.44 (0.10) &  0.44 (0.17) & 0.61 (0.08) \\
\tableline
\multicolumn{5}{c} {NGC4590} \\                         
   HI82  &  -2.44  (0.16) &  0.13  (0.20) &  0.50 (0.30) & 0.50 (0.22) \\
  HI119  &  -2.25  (0.14) &  0.22  (0.20) &  0.70 (0.24) & 0.35 (0.27) \\ 
  HI239  &  -2.03  (0.18) &  0.28  (0.37) & -0.20 (0.35) & 0.30 (0.22) \\
\tableline
average &  -2.26 (0.09)   &  0.19 (0.13)  &  0.44 (0.17) & 0.39 (0.13) \\
\tableline
\multicolumn{5}{c} {M3} \\                              
 265  &  -1.62     (0.14)  &-0.08     (0.16)  &  0.20 (0.25)  & 0.60 (0.17)  \\
 238  &  -2.10     (0.19)  & 0.30     (0.18)  &  0.10 (0.32)  & 0.45 (0.19)  \\
 250  &  -1.40     (0.14)  & 0.38     (0.16)  &  0.50 (0.28)  & 0.40 (0.15)  \\
 334  &  -1.71     (0.14)  & 0.36     (0.16)  &  0.30 (0.22)  & 0.75 (0.15)  \\
 589  &  -1.77     (0.19)  & 0.43     (0.16)  &  0.30 (0.27)  & 0.65 (0.15)  \\
 640  &  -1.51     (0.14)  &-0.02     (0.16)  &  0.30 (0.28)  & 0.55 (0.15)  \\
 885  &  -1.61     (0.15)  & 0.40     (0.17)  &  0.05 (0.34)  & 0.45 (0.16)  \\
1217  &  -1.68     (0.16)  & 0.60     (0.20)  &  0.10 (0.23)  & 0.50 (0.18)  \\
\tableline
average &  -1.64 (0.05)   & 0.28 (0.06)    &  0.24 (0.09)   & 0.55 (0.06)  \\
\tableline
\multicolumn{5}{c} {M107} \\                            
S62  &   -1.03     (0.19)    &  0.55  (0.23) & 0.50 (0.27)    & 0.50 (0.20)  \\
Sf   &   -0.94     (0.18)    &  0.60  (0.23) & 0.15 (0.26)    & 0.40 (0.20)  \\
Sh   &   -1.18     (0.16)    &  0.72  (0.20) & 0.35 (0.24)    & 0.50 (0.17)   \\
Sl   &   -1.17     (0.21)    &  0.68  (0.26) & 0.35 (0.33)    & 0.50 (0.22)   \\
Sr   &   -1.01     (0.20)    &  0.55  (0.24) & 0.25 (0.27)    & 0.35 (0.21)   \\
Ss   &   -1.05     (0.19)    &  0.89  (0.27) & 0.10 (0.29)    & 0.40 (0.20)   \\
Su   &   -0.92     (0.23)    &  0.13  (0.40) &-0.10 (0.33)    & 0.15 (0.24)   \\
\tableline
average &   -1.05 (0.07)    & 0.63 (0.09)    & 0.25 (0.11)    & 0.41 (0.08)  \\
\tableline
\enddata
\end{deluxetable}
\clearpage

\begin{deluxetable}{lrrrr}
\tablewidth{0 pt}
\tablecaption{Derived Abundances for Leo II Sample }
\tablehead{
\colhead{Star } &
\colhead{[FeI/H] ($\sigma$) } &
\colhead{[TiI/FeI] ($\sigma$) } &
\colhead{[MgI/FeI] ($\sigma$) } &
\colhead{[CaII/FeI] ($\sigma$) } 
}
\startdata
 180   & -1.94      (0.18)  &  -0.16     (0.16)   & 	 	    \nodata &0.25  (0.15)    \\
 209   & -1.97      (0.17)  &  -0.13     (0.16)   & 	 -0.20	   (0.30)   &0.30  (0.15)     \\
 271   & -1.77      (0.18)  &  -0.10     (0.18)   & 	0.20	   (0.30)   &0.50  (0.17)     \\
 351   & -1.73      (0.22)  &  -0.17      (0.21)   & 	0.25	   (0.28)   &0.30  (0.18)     \\
 166   & -1.85      (0.18)  &  0.18      (0.19)   & 	0.10	   (0.30)   &0.60  (0.16)     \\
 293   & -1.93      (0.18)  &  -0.05      (0.17)   & 	-0.20	   (0.15)   &0.35  (0.15)     \\
 336   & -1.98      (0.21)  &           \nodata   & 	0.10	   (0.32)   &0.20  (0.17)     \\
 236   & -2.26      (0.22)  &  0.33      (0.21)   & 	0.50	   (0.33)   &0.40  (0.20)     \\
 195   & -1.70      (0.19)  &  0.36      (0.23)   & 	0.70	   (0.30)   &0.42  (0.18)    \\
 258   & -1.64      (0.19)  &  -0.17     (0.19)   & 	0.05	   (0.30)   &0.30  (0.17)     \\
 254   & -1.80      (0.19)  &  -0.10     (0.18)   & 	0.40	   (0.27)   &0.20  (0.17)     \\
 255   & -2.05      (0.19)  &  0.30      (0.17)   & 	-0.35	   (0.39)   &0.45  (0.16)     \\
 296   & -2.13      (0.19)  &  0.28      (0.22)   & 	0.30	   (0.33)   &0.40  (0.17)     \\
 304   & -2.09      (0.18)  &  -0.02      (0.18)   & 	0.45	   (0.27)   &0.35  (0.16)     \\
 281   & -2.07      (0.21)  &  0.35      (0.26)   & 	0.15	   (0.36)   &0.35  (0.20)     \\
 282   & -1.68      (0.19)  &  -0.34     (0.19)   & 	-0.30	   (0.37)   &0.05  (0.17)     \\
 333   & -1.95      (0.21)  &  0.28      (0.21)   & 	0.35	   (0.30)   &0.50  (0.19)     \\
 285   & -1.73      (0.18)  &  -0.03      (0.20)   & 	0.05	   (0.28)   &0.45  (0.16)     \\
 379   & -1.71      (0.21)  &  -0.03     (0.21)   & 	0.15	   (0.33)   &0.30  (0.20)     \\
 260   & -1.75      (0.21)  &  -0.17     (0.20)   & 		   \nodata  &0.35  (0.18)     \\
 341   & -1.42      (0.19)  &  -0.18     (0.19)   & 	-0.00	   (0.30)   &0.15  (0.17)     \\
 256   & -1.73      (0.18)  &   0.05     (0.18)   & 	-0.20	   (0.16)   &0.35  (0.16)     \\
 248   & -1.72      (0.20)  &   0.03     (0.21)   & 	0.05	   (0.18)   &0.40  (0.19)     \\
 420   & -1.80      (0.22)  &  0.24      (0.25)   & 	0.45	   (0.32)   &0.45  (0.21)     \\
 230   & -1.88      (0.24)  &  0.07      (0.25)   & 	0.75	   (0.34)   &0.50  (0.23)     \\
 377   & -1.98      (0.20)  &  0.09      (0.24)   & 	0.15	   (0.38)   &0.40  (0.19)    \\
 234   & -1.76      (0.19)  &  -0.20     (0.21)   & -0.10	   (0.33)   &0.25  (0.18)     \\
\enddata
\end{deluxetable}

\clearpage
\epsscale{0.8}
\plotone{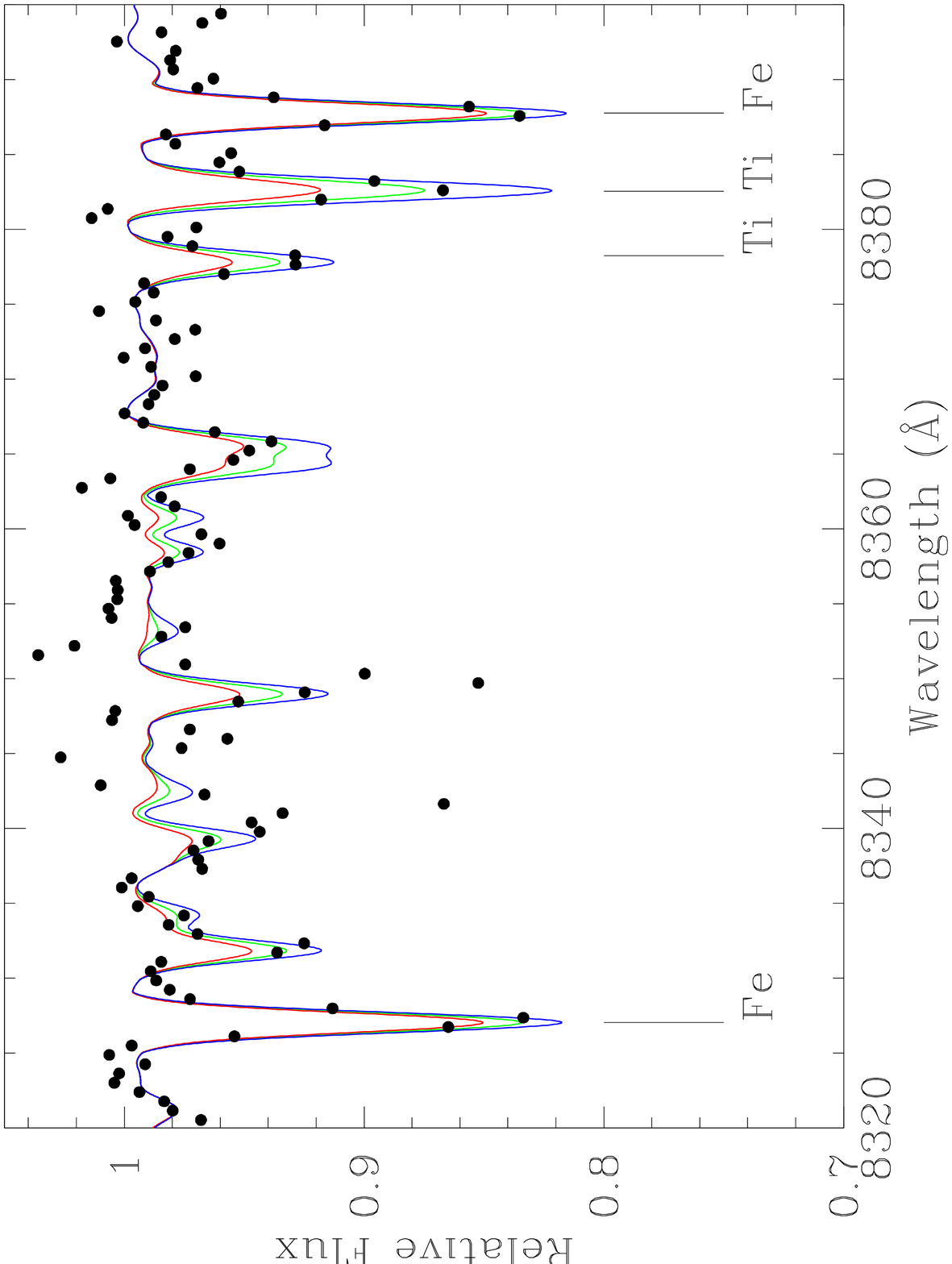}
\figcaption{Modeling the stellar spectra with MOOG.
The points represent the spectrum of Leo II star 180, one of the 
higher SNR spectra.  The lines represent synthetic spectra with 0.3 dex 
higher (blue) and lower (red) Fe and Ti abundances.  
Two Ti lines and two Fe lines used in this analysis are labeled.
\label{fig1}}

\clearpage
\epsscale{0.8}
\plotone{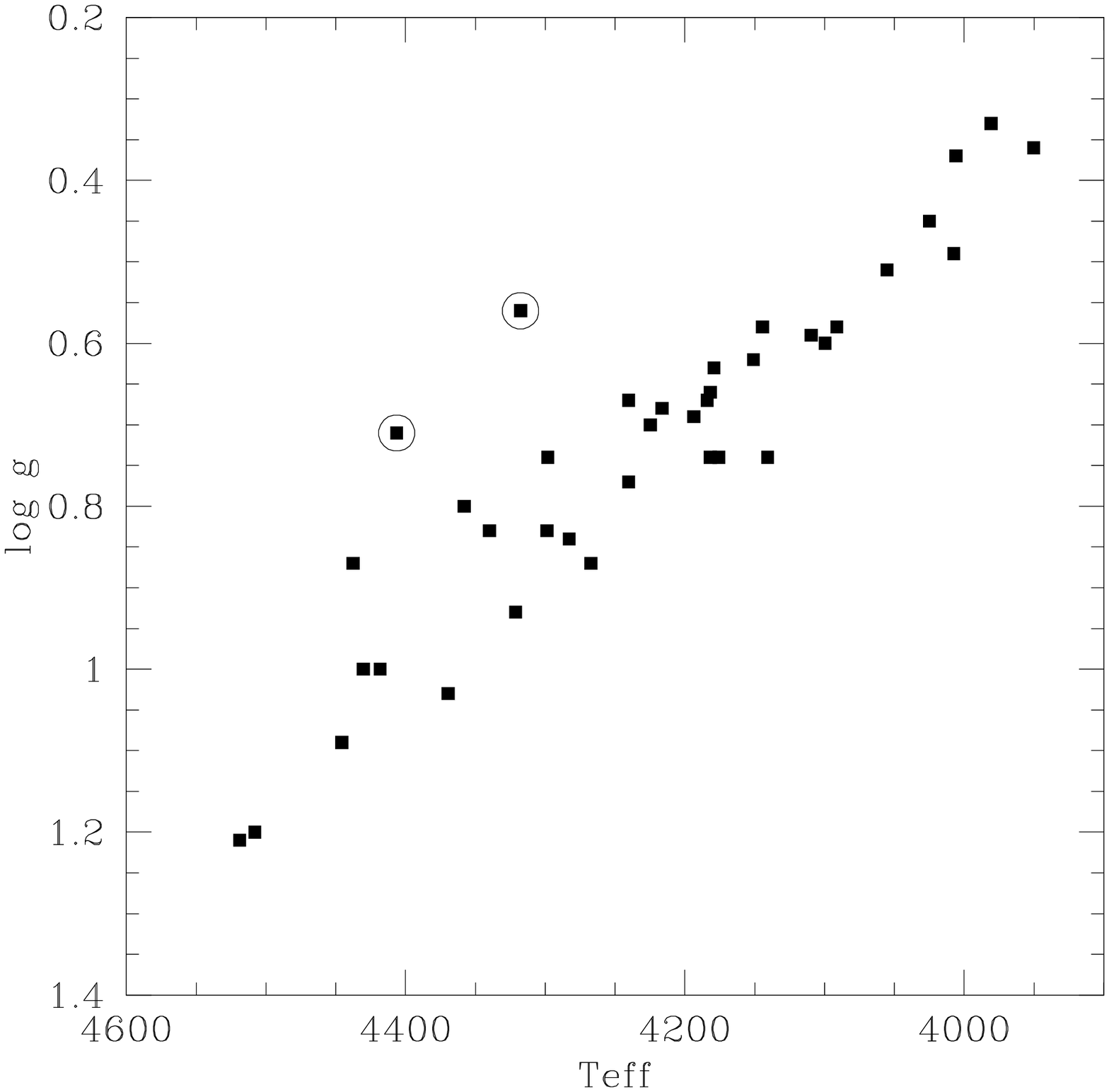}
\figcaption{An HR diagram of the Leo II stars.   The two star with large circles
are stars 166 and 195 which are rv members but don't seem to fall on the 
RGB locus and thus are suspect for our analysis.
\label{fig2}}

\clearpage
\epsscale{0.8}
\plotone{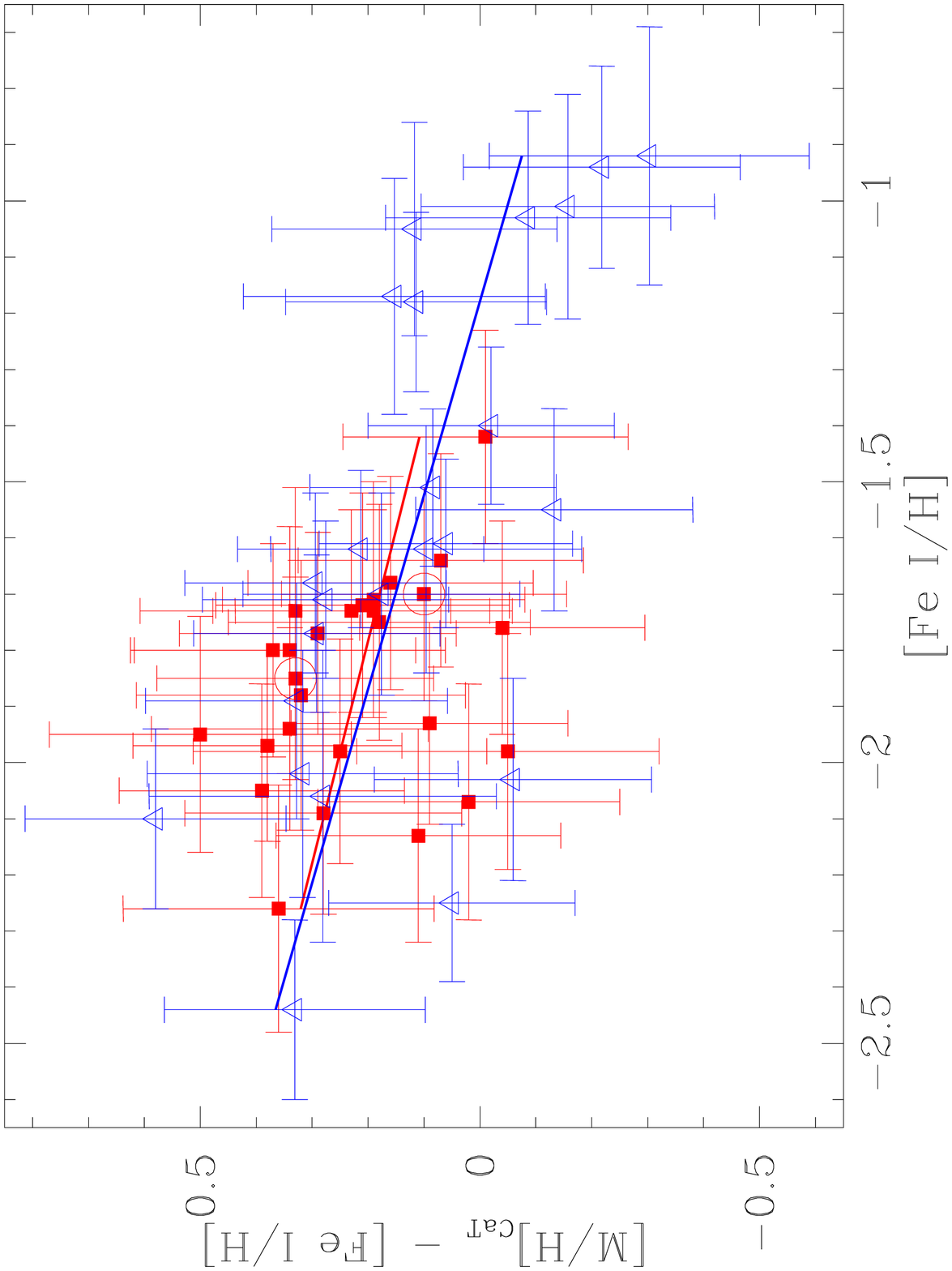}
\figcaption{The [Fe/H] computed from this analysis are compared against the CaT metallicities
for the Leo II stars (red filled squares) and the globular clusters (blue open triangles).   
The thick lines are the best fit lines to the data. The two star with large circles
are stars 166 and 195.
\label{fig3}}

\clearpage
\epsscale{0.8}
\plotone{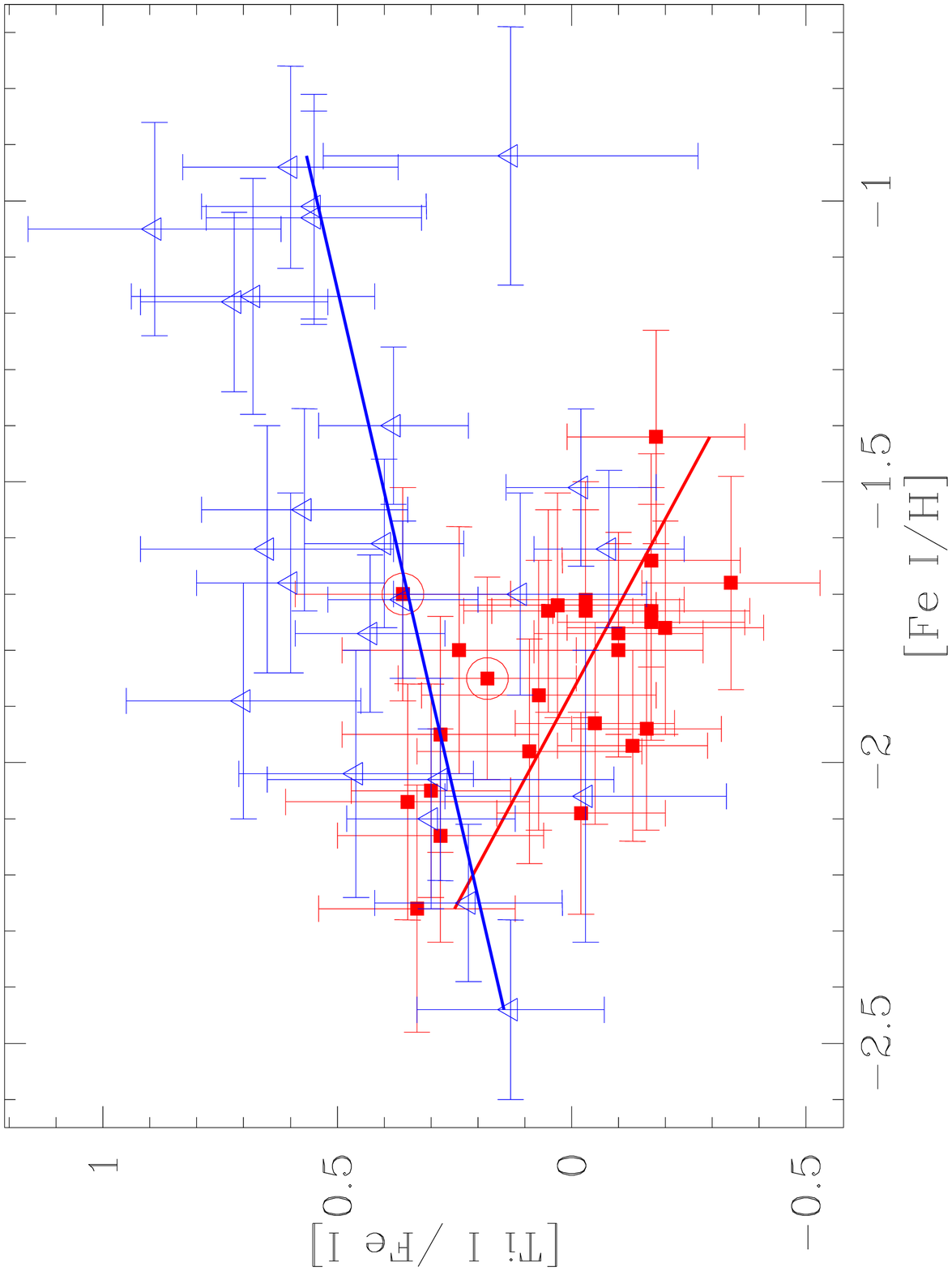}
\figcaption{The [Ti/Fe] computed from this analysis is shown 
for the Leo II stars (red filled squares) and the globular clusters (blue open triangles).   
The thick lines are the best fit lines to the data. The two star with large circles
are stars 166 and 195.
\label{fig4}}

\clearpage
\epsscale{0.8}
\plotone{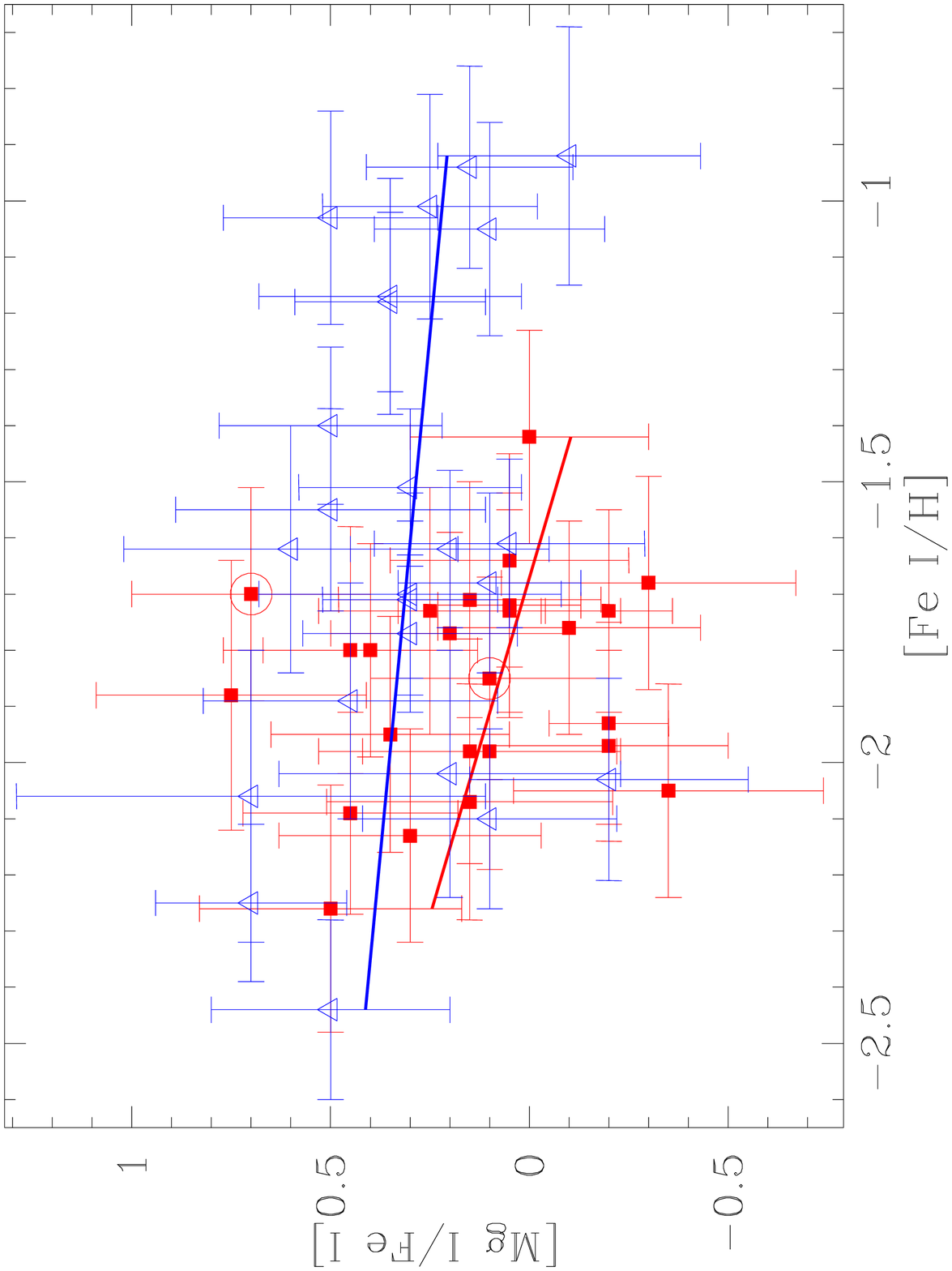}
\figcaption{The [Mg/Fe] computed from this analysis is shown 
for the Leo II stars (red filled squares) and the globular clusters (blue open triangles).   
The thick lines are the best fit lines to the data. The two star with large circles
are stars 166 and 195.
\label{fig5}}

\clearpage
\epsscale{0.8}
\plotone{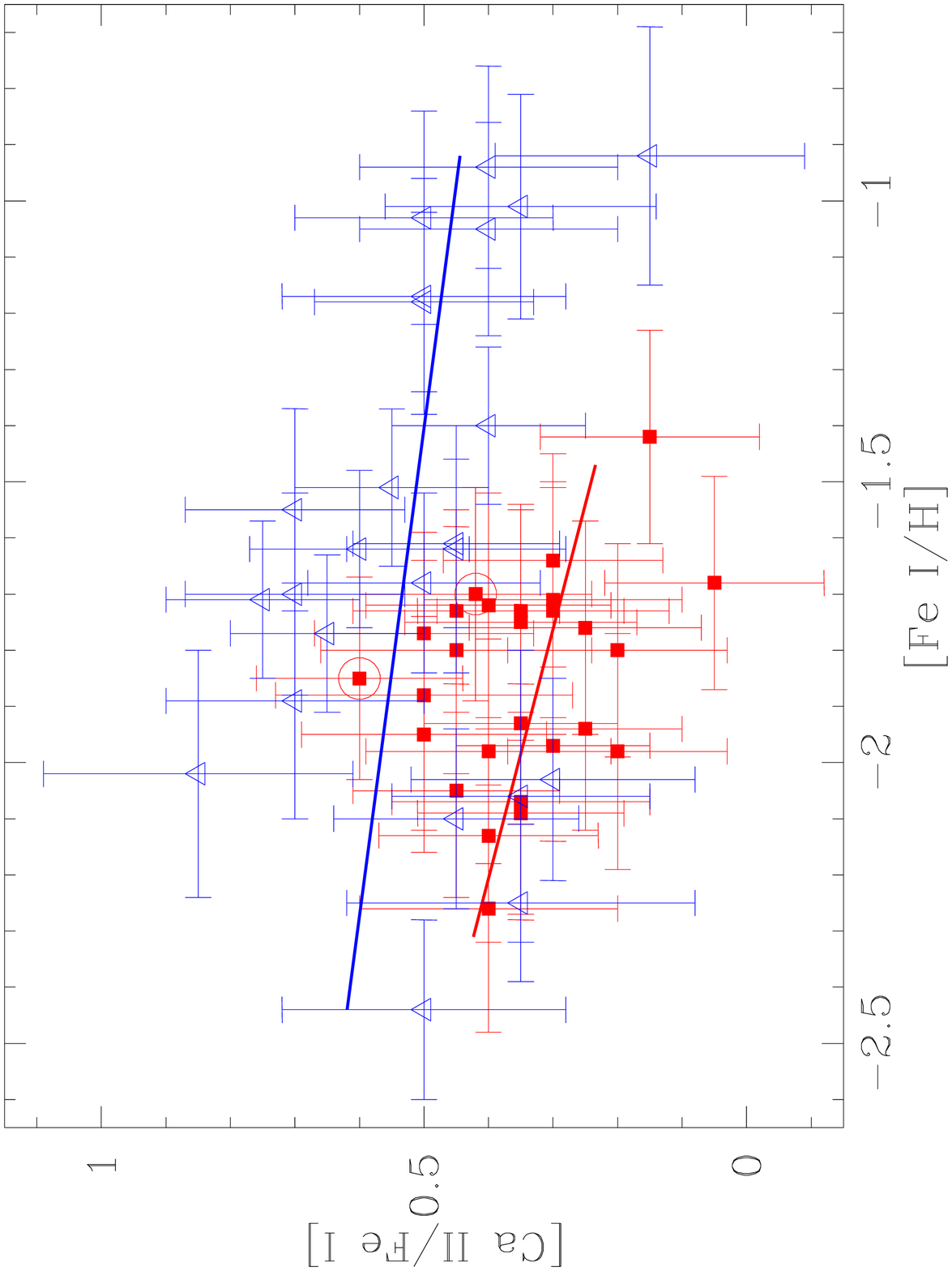}
\figcaption{The [Ca II/Fe I] computed from this analysis is shown 
for the Leo II stars (red filled squares) and the globular clusters (blue open triangles).   
The thick lines are the best fit lines to the data. The two star with large circles
are stars 166 and 195.   Please see the Section 3 and Figure 7 for a full discussion 
of the errors and zero points associated with the Ca abundances shown in this figure.
\label{fig6}}

\clearpage
\epsscale{0.8}
\plotone{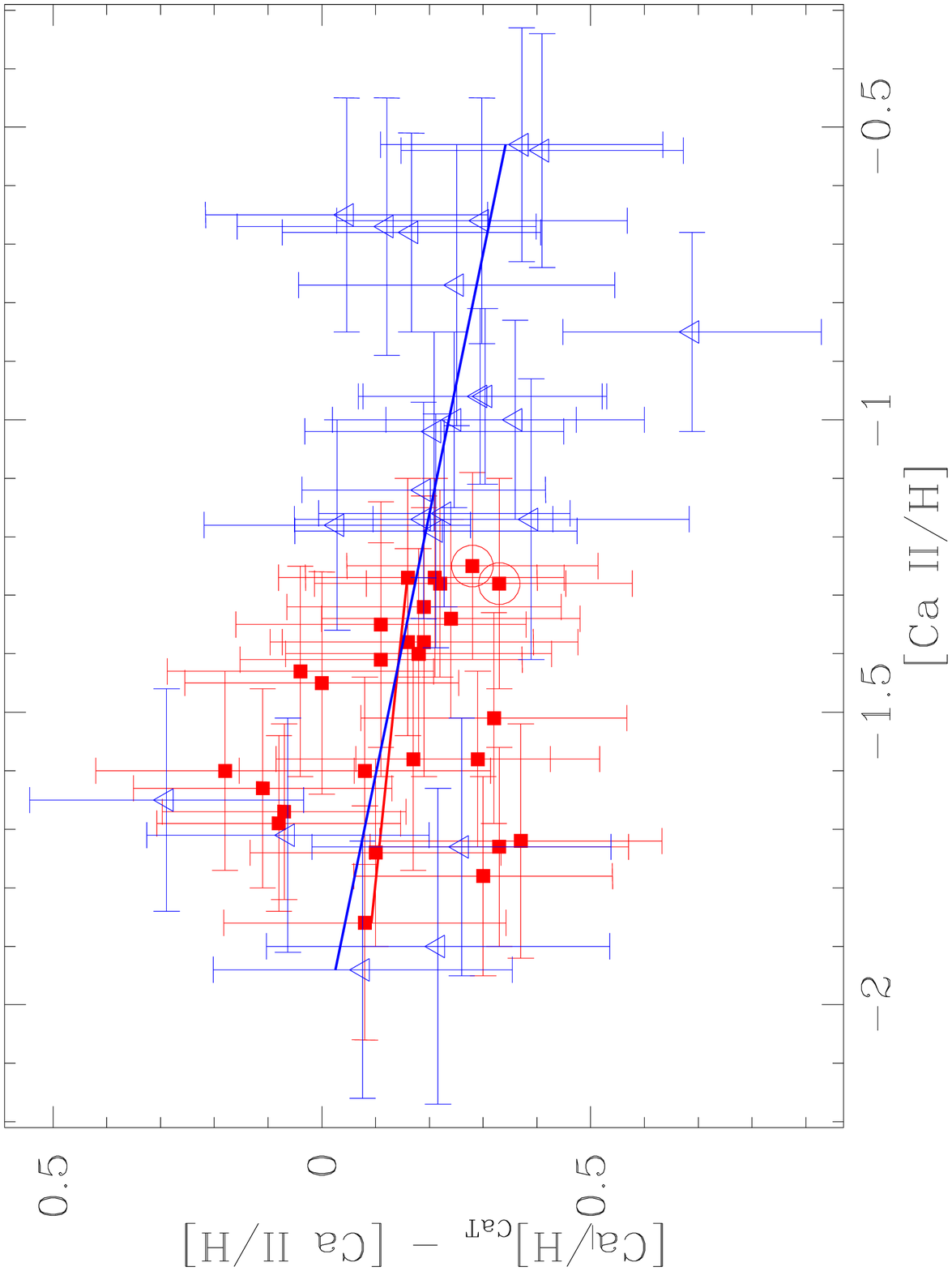}
\figcaption{The [Ca/H] computed from this analysis are compared against the CaT
[Ca/H] metallicities
for the Leo II stars (red filled squares) and the globular clusters 
(blue open triangles) from BSS07.
The thick lines are the best fit lines to the data. The two star with large 
circles are stars 166 and 195.
\label{fig7}}

\clearpage
\epsscale{0.8}
\plotone{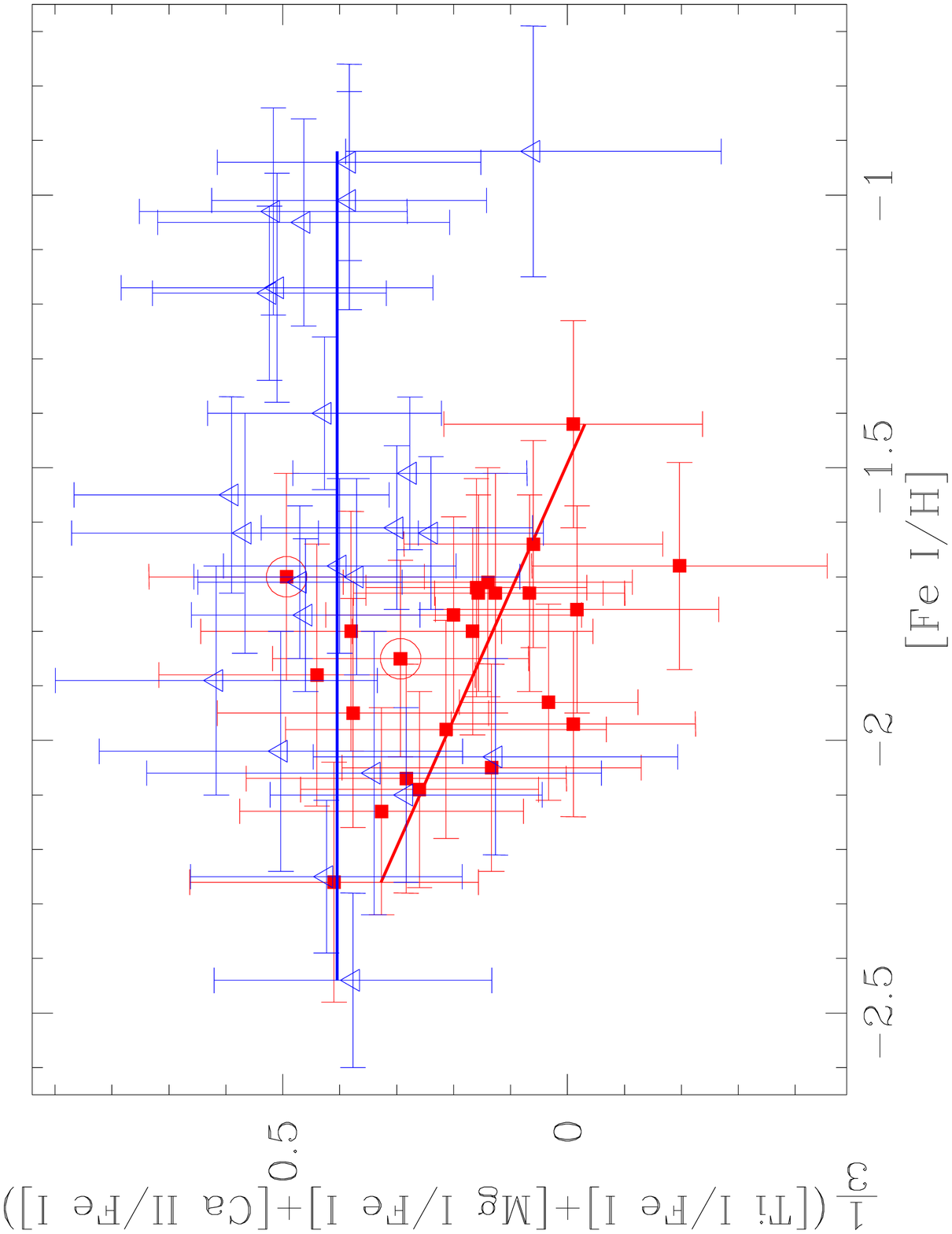}
\figcaption{The average $\alpha$-abundance for the globular cluster (open blue triangles) 
and Leo II samples (red filled squares) are shown from the unweighted Ca, Mg and Ti abundance
ratios.  The thick lines are the best fit fits to the data.
The two star with large circles are stars 166 and 195 and are not included in the best fit.
\label{fig8}}

\clearpage
\epsscale{0.8}
\plotone{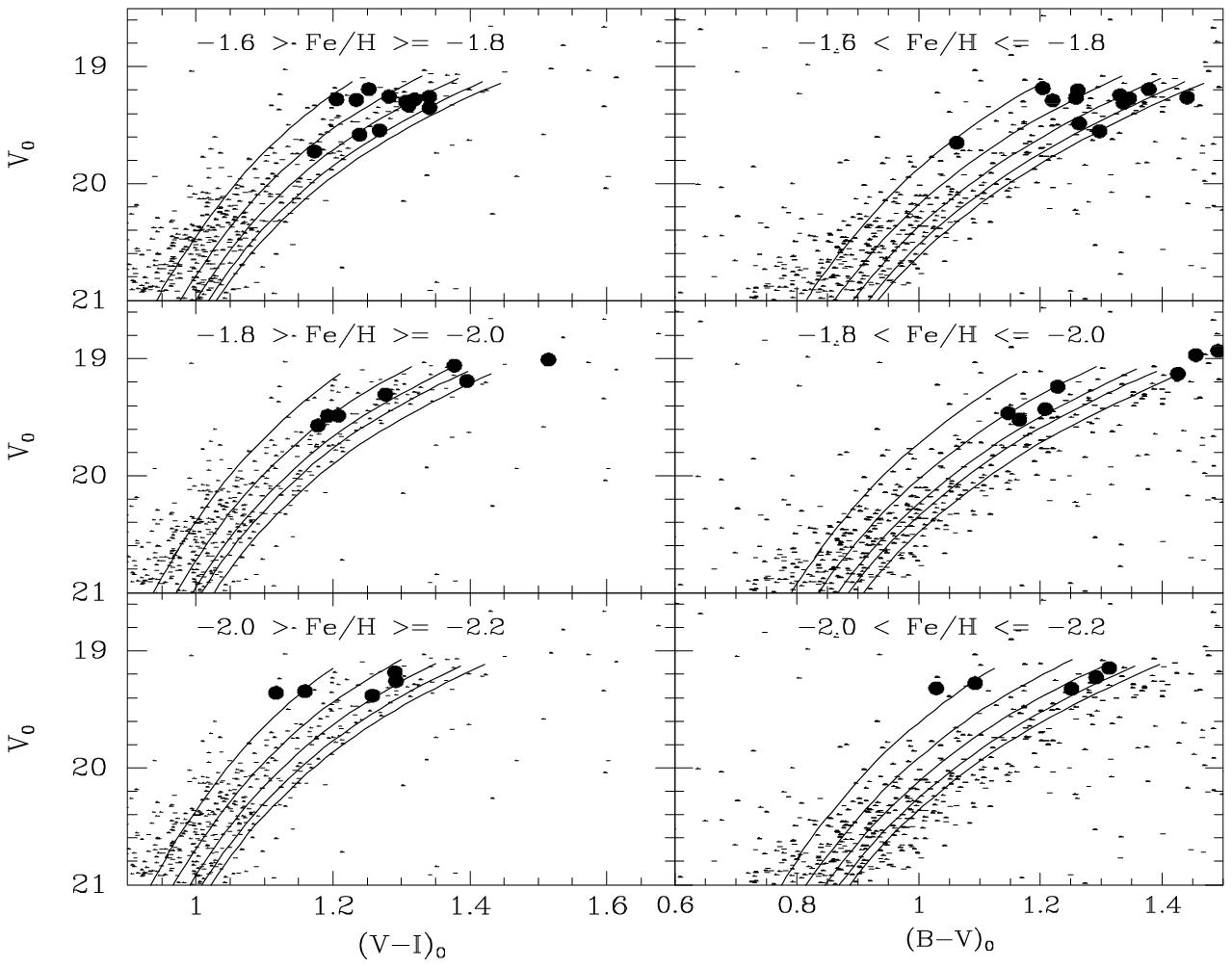}
\figcaption{Calculation of approximate Leo II RGB star ages.  The panels are broken
down by metallicity and passband.  The isochrones correspond, from bluest to reddest, to 
age of 3, 6, 9, 12 and 15 Gyr.  Note the two outlying stars in the bottom panels
which are likely AGB stars.  These are different stars from the two outliers highlighted
in figures 3-8, which have been excluded {\it a priori}.
\label{fig9}}

\clearpage
\epsscale{0.8}
\plotone{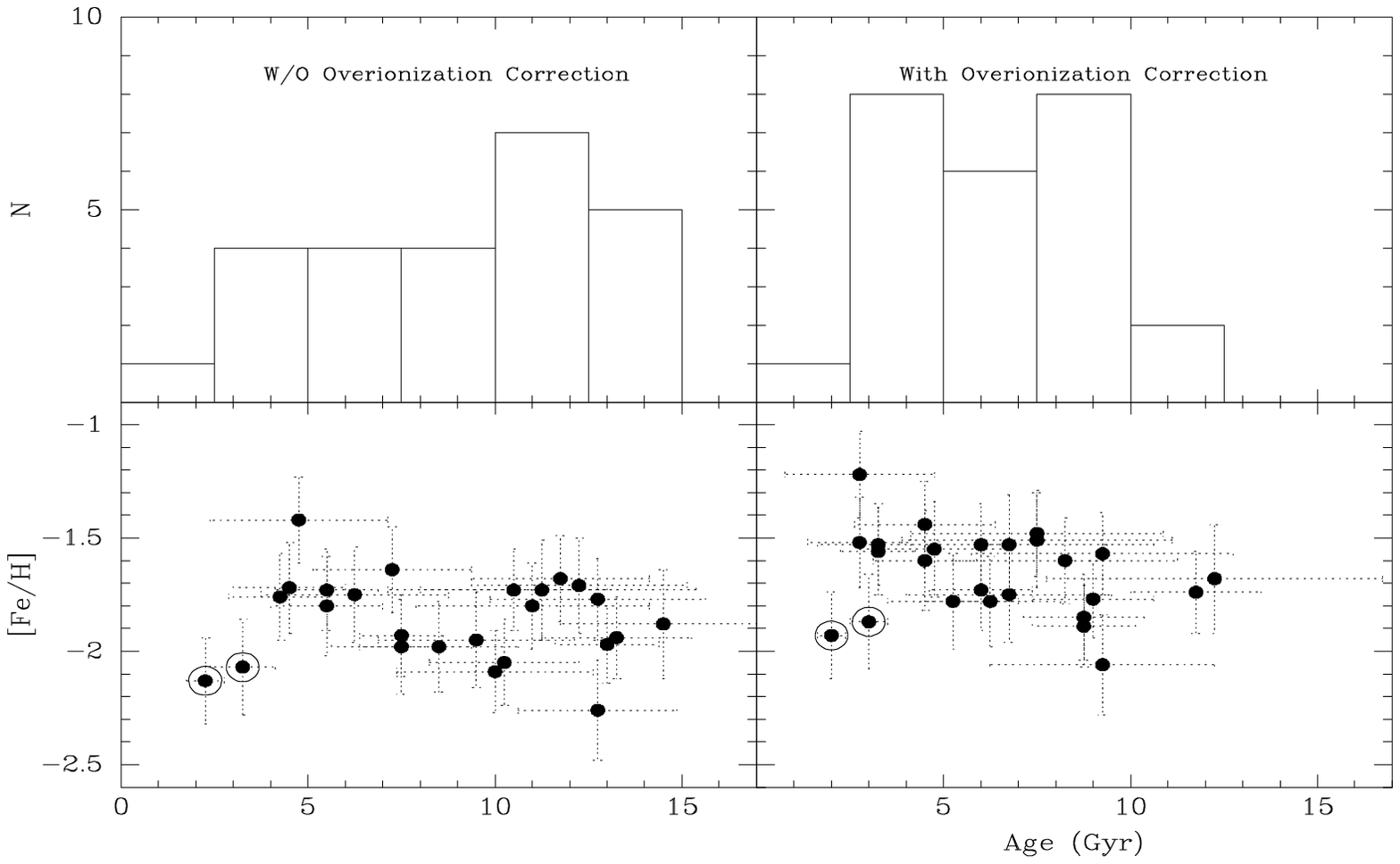}
\figcaption{The age distribution and age-metallicity relationship for the
RGB stars of Leo II.  The top panels show a histogram of the inferred ages while
the lower panels show the AMR.  The left panels shows the raw distributions while
the right panels are corrected for
overionization of the Fe\,I lines. The circled points are the outlier stars, which
are different stars than the two outliers rejected in figures 3-8.
\label{fig10}}

\end{document}